# Anisotropic Charge Carrier and Coherent Acoustic Phonon Dynamics of Black Phosphorus Studied by Transient Absorption Microscopy

Shengjie Meng[1], Hongyan Shi[1,2], Hu Jiang[1], Xiudong Sun[1,2], Bo Gao[1,2*]

[1]Institute of Modern Optics, School of Physics, Key Laboratory of Micro-Nano Optoelectronic Information System, Ministry of Industry and Information Technology, Key Laboratory of Micro-Optics and Photonic Technology of Heilongjiang Province, Harbin Institute of Technology, Harbin 150001, China

[2]Collaborative Innovation Center of Extreme Optics, Shanxi University, Taiyuan 03006, China

KEYWORDS: Anisotropic Charge Carrier Dynamics, Anisotropic Coherent Acoustic Phonon Dynamics, Black Phosphorus, Transient Absorption Microscopy, Acoustic Echo

**Abstract**

Due to its corrugated hexagonal lattice structure, Black phosphorus (BP) has unique anisotropic physical properties, which provides an additional freedom for designing devices. Many body interactions, including interactions with phonon, is crucial for heat dissipation and charge carrier mobility in device. However, the rich properties of the coherent acoustic phonon, including anisotropy, propagation and generation were not fully interrogated. In this paper, the polarization-resolved transient absorption microscopy was conducted on BP flakes to study the dynamics of photoexcited charge carriers and coherent acoustic phonon. Polarization-resolved transient absorption images and traces were recorded and showed anisotropic and thickness-dependent charge carriers decay dynamics. The damping of the coherent acoustic phonon oscillation was found to be anisotropic, which was attributed to the polarization-dependent absorption length of the probe pulse. From the analysis of initial oscillation amplitude and phase of coherent acoustic phonon oscillation, we proposed that the direct deformation potential mechanism dominated the generation of coherent acoustic phonons in our experiment. Besides, we obtained the sound velocity of the coherent acoustic phonon from the oscillation frequency and the "acoustic echo", respectively, which agreed well with each other. These findings provide significant insights into the rich acoustic phonon properties of BP, and promise important application for BP in polarization-sensitive optical and optoelectronic devices.

**Introduction**

Black phosphorus (BP) is a promising two-dimensional semiconductor material with thickness dependent direct bandgap from 0.3 (bulk)[1] to ~2 eV (single layer)[2] and high carrier mobility (on order of $10^4$ $cm^2V^{-1}s^{-1}$),[3-5] which endows BP a promising candidate for further applications such as electronic device with high speed and on-off current ratios,[6-10] mid-infrared optoelectronic device with lower noise.[11-15] Besides, its corrugated hexagonal lattice structure provides unique anisotropic physical properties, such as optical responses,[16] carrier mobility[4-5] and thermal conductivity,[17-18] which brings a new direction in designing novel and polarization sensitive electronic, optoelectronic and optomechanical devices.[5, 13, 15, 19-22]

A lot of angle-resolved optical techniques have been used to study the anisotropic response of BP, such as Raman spectroscopy,[19, 23-24] photoluminescence spectroscopy[19, 25] and scanning polarization modulation microscopy.[26-27] But due to far lower vibrational energy than optical phonons, acoustic phonons were difficult to be detected by traditional Raman spectroscopy.[28] Transient absorption method based on pump-probe technique provides a good choice for observing coherent acoustic phonons in solids when the probe energy resonates with the transient absorption.[29-31] By transient absorption method, ultrafast charge carrier dynamics of BP has been investigated.[32-36] These ultrafast studies showed anisotropic charge carrier dynamics.[32, 34, 36] For designing better device based on these unique anisotropic properties, it is far from being enough just making clear the behavior of charge carriers. In a pump-probe based ultrafast process, the pump pulse excitation will also set up a transient stress and generate a strain wave, which will propagate from the sample surface at the speed of longitudinal acoustic phonon. The strain wave bounces back and forth inside the sample and caused a periodic modulation of the local dielectric constants. When the time-delayed probe pulse is incident to the surface of the sample, the absorption of the probe pulse will change. As a result, the reflectivity of the time-delayed probe pulse is modified by the generation of coherent longitudinal acoustic phonons.[37-41] It is remarkably necessary to elucidate the many body interactions in BP, including interactions with the phonon, which were crucial for heat dissipation and charge carrier mobility in device.[42-43] One recent study revealed layer-dependent coherent acoustic phonon behaviors of BP using transient absorption microscopy.[29] However, the anisotropy of the coherent acoustic phonon dynamics was not fully interrogated.

In this paper, we investigated the anisotropic dynamics of photoexcited charge carriers and coherent acoustic phonon in BP using polarization-resolved transient absorption microscopy. Polarization-resolved transient absorption images and traces were recorded and showed anisotropic and thickness-dependent charge carriers decay dynamics. The damping time of coherent acoustic phonon oscillation was also found to be dependent on the probe polarization, which was attributed to polarization-dependent absorption length of the probe pulse. By fitting the transient absorption traces for beam polarization along armchair and zigzag directions, we found that the initial oscillation amplitude of coherent acoustic phonon was anisotropic and had a linear relationship with pump power, from which and phase analysis we believed that the direct deformation potential mechanism dominated the generation of coherent acoustic phonons in our experiment. Besides, we calculated the sound velocity of the coherent acoustic phonon from the oscillation frequency and the "acoustic echo", respectively, which agreed well with each other.

**Methods**

As shown in Figure 1a, BP is a layered material with orthorhombic crystal structure and its corrugated hexagonal lattice structure result in its unique in-plane anisotropy.[4-5, 16-18,

[44] Top view of single layer BP was shown in Figure 1b, in which the armchair and zigzag direction were labelled. Band diagram of BP was shown in Figure 1c, in which pump (green arrow) and probe (red arrow) photon transition were labelled. For better investigating the transient absorption dynamics with coherent acoustic phonon oscillation in BP, we built a transient absorption microscope with temporal resolution of ~600 fs and spatial resolution of ~200 nm.

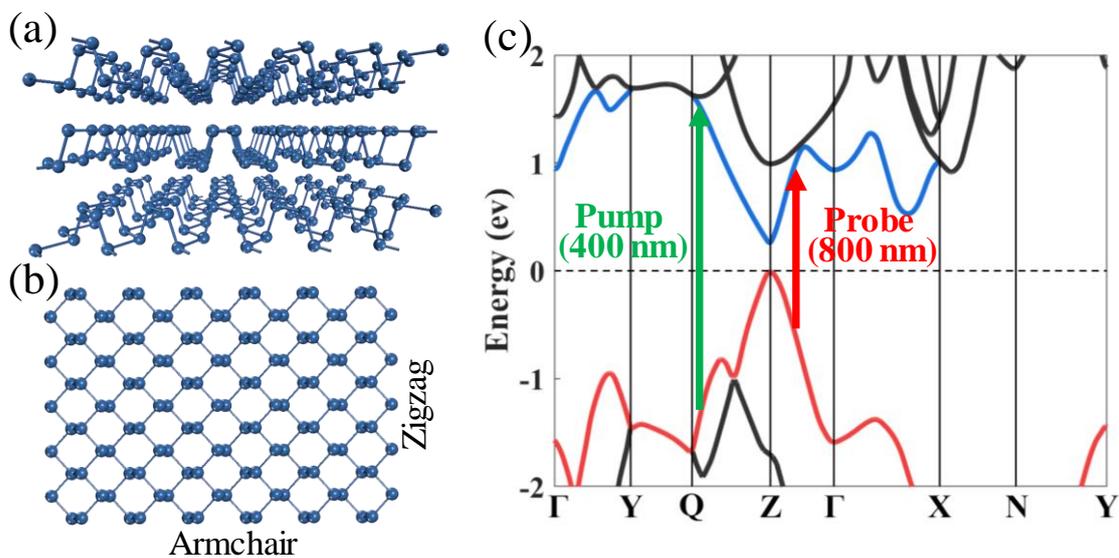

Figure 1. Schematic diagram for geometric structure of BP. (a) Side view of multilayer BP. (b) Top view of single layer BP, in which the armchair and zigzag direction were labelled. (c) Band structure of BP, in which pump (green arrow) and probe (red arrow) photon induced transitions were labelled.

The schematic diagram of home-built transient absorption microscope was shown in Figure 2. A Ti: Sapphire oscillator (Micra, Coherent) provides fs laser pulse at repetition rate of 80 MHz with a wavelength of 800 nm, which was split into two beams by a beam splitter. One of the beams (70% of the total energy) was doubled by a $\beta$-barium borate (BBO) crystal and used as the pump pulse, while the other beam (30% of the total energy) served as the probe pulse. An acousto-optic modulator (R23080-1-1.85-LTD, Gooch &

Housego) modulated the pump beam at 100 kHz. By combination of a polarizer and a half-wave plate, the polarization direction of the pump and probe beams was made parallel or perpendicular to each other. Then, the two beams were spatially overlapped and conducted into an inverted microscope (Ti-U, Nikon) equipped with a 60×, 1.49 numerical aperture (NA) oil-immersion objective and a piezo stage (Nano-View/M-100-3, MadCity Lab) which was used for scanning. By the high NA objective, we could get a small laser spot with spatial resolution of ~200 nm. The same objective was used to collect the reflected beams. After filtering the pump beam, only the reflected probe beam was detected by an avalanche photodiode (APD120A/M, Thorlabs). The reflectivity changes ($\Delta R$) of the probe beam due to the pump beam was monitored with a lock-in amplifier (SR830, Stanford Research Systems). Finally, the signal was input into the computer for further processing. Using this setup, we can do localized excited states study on BP flakes with low shot-noise and high sensitivity, and perform excited states imaging of BP flakes.

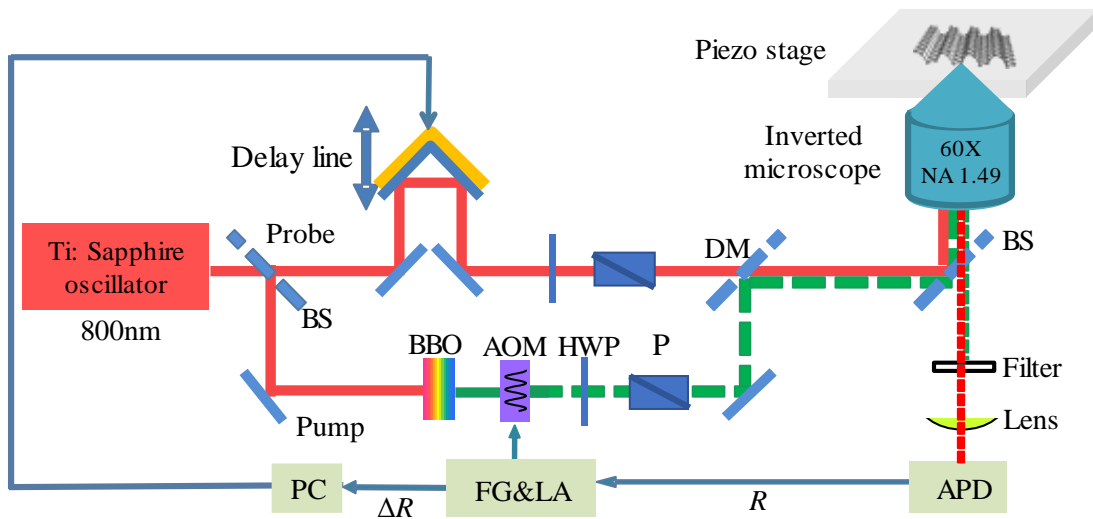

Figure 2. Schematic diagram for home-built transient absorption microscopy. BS: beam splitter; BBO: *β*-barium borate; AOM: acousto-optic modulator; HWP: Half-wave plate; P: polarizer; DM: dichroic mirror; APD: avalanche photodiode; FG: function generator; LA: lock-in amplifier; PC: personal computer.

**Results and Discussion**

Thin BP fakes on glass substrate were prepared by standard mechanical exfoliation from bulk BP. Subsequently, freshly exfoliated thin BP flakes were used directly for characterization and optical measurements at temperature of ~21 °C and relative humidity of <10%. In our experiment, before the Raman spectra, AFM and transient absorption measurements, we would carefully adjust the orientation of the sample to make sure that the BP flakes were oriented in the same way to improve the experimental accuracy. Comparing the images obtained from optical microscope, AFM and transient absorption microscope (shown in Figs. 3a-3c), we could see that they overlap quite well. Figure 3a shows a typical optical image of thin BP flake with size in tens of microns. Angle-resolved polarized Raman spectra (ARPRS) were collected to determine the crystalline orientation by exciting the BP flake with 532 nm laser line. From the polarized Raman spectra of the BP flake excited at different polarization angles and the polar plot of the integrated intensity of $A_g^2$ mode (see Figure S1 in Supporting Information), we could obtain that the zigzag direction is along the black line in Figure 3a, which has a cross angle of 6° with horizontal direction rotated anticlockwise.[45-46] Atomic force microscopy (AFM) image in Figure 3b shows that the BP flake is mainly composed of a few flat terraces, with thickness of two large terraces labelled: region 1 for 500 nm and region 2 for 392 nm. To obtain rich spatial-temporal information of the BP flake, the piezo stage was used for scanning the sample at fixed pump-probe delays.

Figure 3c shows the transient absorption image of the BP flake recorded at pump-probe delay 0 ps with both pump and probe beams polarized along zigzag direction. It can be seen that the transient absorption signals showed an obvious thickness dependence. E.g., at region 1 the transient absorption signal is positive, while at region 2 it is negative. Similar results could also be seen in transient absorption images recorded at pump-probe delay of 10 ps and images obtained with other beam polarizations (see Figure S2 in Supporting Information). The difference could be more clearly seen in decay dynamics (shown in Figure 3d), in which transient absorption traces were compared for the BP flake

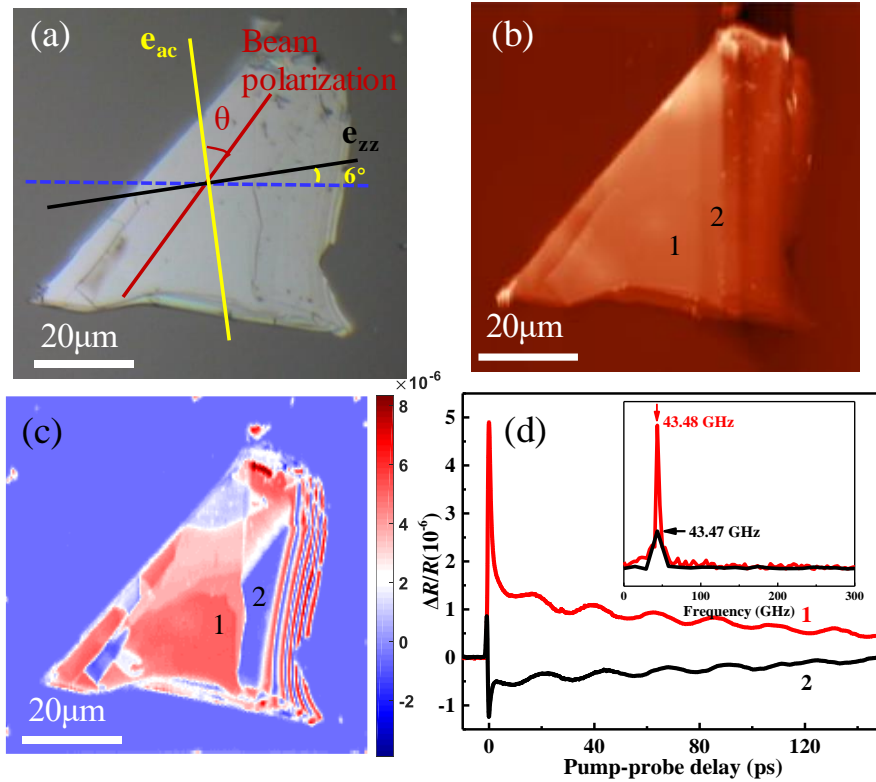

Figure 3. (a) Optical image of one BP flake on glass slide. Crystallographic directions of BP are defined as follows: $e_{ac}$ is the armchair direction and $e_{zz}$ is the zigzag direction. The red solid line is the polarization direction of the incident beam. $\theta$ is the angle of polarization direction measured clockwise from armchair direction. The blue dashed line is horizontal direction in the experiment. (b) Atomic force microscopy image of the BP flake in (a). The thickness of two regions labelled by 1 and 2 are 500 nm and 392 nm, respectively. (c) Transient absorption image of the BP flake taken at pump-probe delay of 0 ps with both beams polarization along zigzag direction. (d) Transient absorption traces recorded at region 1 and region 2 with both beams polarization along zigzag direction. The inset shows the fast Fourier transform of the oscillations in the two transient absorption traces. The pump and probe power were 90 μW and 260 μW, respectively.

at region 1and region 2. At region 1, BP exhibit positive transient reflectivity (Δ*R*/*R*) at 0 ps pump-probe delay, meaning more reflection of the probe with presence of the pump, which corresponds to a photoinduced bleach (PB) signal. At region 2, BP exhibited negative Δ*R*/*R* at 0 ps pump-probe delay, corresponding to a photoinduced absorption (PA) signal. Subsequently, Δ*R*/*R* decayed slowly with a long lifetime. The thickness-dependent transient absorption dynamics were also seen at other thicknesses (see Figure S3 in Supporting Information). The BP flakes used in our experiment are very thick and could be thought to have identical band structures. Therefore, the difference in transient absorption dynamics of BP flakes at different thicknesses was not originated from different band structures. We speculate that the thickness-dependent transient absorption dynamics is related to multi-reflection.[27, 45-46]

In addition to the usual photoinduced charge carrier relaxation, a strongly coherent acoustic phonon oscillation was clearly observed in the traces as the probe laser energy 1.55 eV was in resonance with the transient absorption.[29] Meantime, the oscillation amplitude was gradually damped, which could be attributed to three factors[47-49]: one is the absorption induced attenuation of the strain wave when it propagated from the surface to the inner part of the BP flake; the second is the acoustic phonon decay; the third is the fast radiative energy loss from the BP flake to the surroundings. Fast Fourier transform of the two traces shown in the inset of Figure 3d give the oscillation frequency $f$ = 43.48 and 43.47 GHz for thickness at region 1 and 2, respectively. Unrelated with the thickness, the oscillation frequency has a relation with sound velocity $v$ in the form of $1/f = \lambda/2nv$, where $\lambda$ = 800 nm is the wavelength of the probe, $n$ = 3.26 is the refractive index of BP in the zigzag direction.[26, 50-51] Then, we obtained the sound velocity $v = 5.33 \times 10^3$ m/s, which is well consistent with previous studies.[52-56] Furthermore, by the equation $Y = \rho v^2$, in which $\rho = 2.69 \times 10^3$ kg/m³ is the density of BP,[57] the Young's modulus $Y$ of the BP flake is calculated to be 76.42 GPa, which is roughly comparable with the value earlier reported.[58]

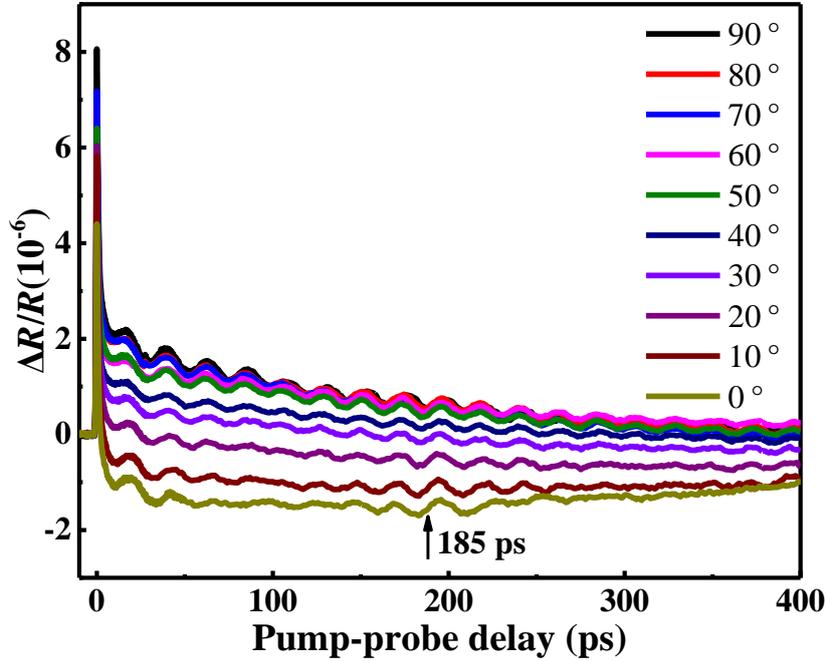

Figure 4. Transient absorption traces recorded at a series of beam polarization from 0° (armchair direction) to 90° (zigzag direction), in which the pump pulse and probe pulse polarization direction parallel with each other. The beam polarization angle is shown in Figure 3a. The pump and probe power were 90 μW and 260 μW, respectively.

It is well known that BP exhibits distinct optical anisotropy due to its corrugated hexagonal lattice.[5] Therefore, the polarization-resolved measurement of the transient absorption dynamics in BP is meaningful. For the BP flake shown in Figure 3a, the transient absorption dynamics traces were measured at region 1 for a series of beam polarization angles $\theta$ from 0° (armchair direction) to 90° (zigzag direction). The polarization angle of the pump pulse and probe pulse were changed by a rotatable half wave plate and made sure that they were parallel to each other, so that we could obtain the anisotropic dynamics response of the BP flake from armchair to zigzag direction. As shown in Figure 4, when $\theta$ was 90°, the BP flake showed strong positive PB signal and an abrupt decrease. The positive PB signal was due to pump pulse induced ground state depletion. Subsequently, the positive PB signal decayed slowly with a long lifetime. When the laser polarization was getting closer to the armchair (0°) direction, the initial signal was positive PB signal, but rapidly decreased and switched to negative photoinduced absorption (PA) signal. We attributed the negative PA signal to the phonon-assisted intraband (Drude) absorption, which gave rise to excited states absorption counteracting pure transient bleaching of interband ground state absorption, as reported in graphene.[59] Subsequently, the negative PA signal decayed slowly with a long lifetime. The gradual change of the transient absorption signal from PB to PA was attributed to the varying carrier density of BP, which further leads to different weights of the intraband absorption depending on the probe polarization. Besides, the coherent acoustic phonon oscillation damped more and more slowly, when the beam polarization was gradually changed from armchair to zigzag direction.

To elucidate whether the change was originated from the pump or the probe beam, we collected the dynamics response of another BP flake using parallel and perpendicular polarization configurations, as shown in Figure S4 in Supporting Information. It can be seen that the coherent acoustic phonon oscillation had little relations to the pump polarization. It was because carrier-carrier scattering randomized their momentum very fast and was beyond the instrument time resolution,[35] although the carriers have the momentum along the polarization direction of the pump after the pump excitation. Meantime, the coherent acoustic phonon oscillation changed with the probe polarization, which would be discussed below.

To obtain the many body interactions time, the transient absorption traces for beam polarization along armchair and zigzag direction were fitted by three exponential decays superposed on one damped oscillation term:

$$y = \sum_{n=1,2,3} A_n \exp\left(\frac{x}{t_n}\right) + B\exp\left(\frac{x}{t_c}\right)\sin(2\pi w_c x - \alpha)$$

where $A_n$ is proportionality constants, $t_1$, $t_2$ and $t_3$ are the lifetime of three photoinduced carrier relaxation processes in BP flake. The first process was attributed to the energy relaxation of hot carriers though carrier-carrier and carrier-phonon scattering after photoexcitation,[35] which was very fast and formed a thermalized Fermi-Dirac distribution. The second process was attributed to the carrier recombination and lattice heating.[35] The third process was attributed to the cooling of lattice temperature through dissipation of heat to the substrate.[34] $B$ is the initial amplitude of the coherent acoustic phonon oscillation. $w_c$ and $t_c$ are the periodic oscillation frequency and lifetime of the damped coherent acoustic phonon oscillation. $\alpha$ is phase of the periodic oscillation.

The fitting results are shown in Table 1 and Figure S5 in Supporting Information. $t_1$ for beam polarization along armchair and zigzag direction are 0.68 ps and 0.65 ps, respectively, which are very fast and roughly the same. Therefore, there are no obvious anisotropic effects in the carrier-carrier and carrier-phonon interactions. $t_2$ for beam polarization along armchair and zigzag direction is 15.83 ps and 5.92 ps, respectively, meaning the carrier recombination and lattice heating is more quickly when the beam polarization gradually changed from armchair direction to zigzag direction. In order to clearly understand the difference of $t_2$ in 0° (armchair direction) and 90° (zigzag direction), we fitting the $t_1$, $t_2$, $t_3$ at different pump power, as shown in Figure 5a and Table S1 in Supporting Information. From the fitting result, we could see that $t_2$ increased with the pump power increasing. Higher pump power led to higher carrier density in the same situation. So we tentatively attributed the slower process to the higher carrier density in the armchair direction caused by the larger absorption coefficient in this direction.[38, 50] $t_3$ for beam polarization along armchair and zigzag direction are 165.32 ps and 154.95 ps, respectively, which are very slow and roughly the same. It is reasonable since heat conductance is a very slow process and the heat dissipation to the substrate is independent of the beam polarization.

Table 1. Fitting results of transient absorption traces for beam polarization along armchair and zigzag direction

| $\theta$ | $t_1$ (ps) | $t_2$ (ps) | $t_3$ (ps) | $t_c$ (ps) | $\omega_c$ (GHz) | $\alpha$(degree) |
|---|---|---|---|---|---|---|
| 0° (armchair) | 0.68(86%) | 15.83(13%) | 165.32(1%) | 42.36(100%) | 42.78 | 48.32 |
| 90° (zigzag) | 0.65(76%) | 5.92(9%) | 154.95(15%) | 123.26(100%) | 43.85 | 16.14 |

For the coherent acoustic phonon oscillation, as shown in Table 1, the damping time $t_c$ for probe polarization along armchair direction is 42.36 ps, much smaller than that along zigzag direction, 123.26 ps, indicating anisotropic damping time. It is known that damping time $t_c$ has a relation with absorption length $\zeta$ in the form of $t_c = \zeta/v$, where $v$ is the sound velocity, $5.33\times10^3$ m/s, because the probe beam is sensitive to the coherent acoustic phonon oscillation just within the absorption depth from the sample surface.[37] The absorption length $\zeta$ could be calculated through the equation $\zeta = c/2\omega\kappa = \lambda/4\pi\kappa$,[38] in which $c$ is the speed of light, $\kappa$ is the imaginary part of refractive index in BP. The refractive index of BP at wavelength of 800 nm is 3.19-0.29i and 3.26-0.1i for beams polarized along armchair and zigzag direction,[50] respectively, showing an anisotropy in both real ($n$) and imaginary ($\kappa$) parts. We could obtain absorption length $\zeta = 218$ nm and 636 nm for the probe polarized along armchair and zigzag direction, respectively, and the damping time $t_c = 40.91$ ps and 119.32 ps for beam polarization along armchair and zigzag direction, respectively, which was consistent with the fitting result in Table 1. Therefore, we attributed the anisotropic damping time to the anisotropic absorption length arising from polarization-dependent refractive index.

As shown in Table 1, the coherent acoustic phonon oscillation frequencies for beam polarization along armchair and zigzag direction are 42.78 GHz and 43.85 GHz, respectively. It is known that the oscillation frequency has a relation with sound velocity $v$ in the form of $1/f = \lambda/2nv$, where $\lambda$ is the wavelength of the probe pulse, $v$ is the sound velocity, $n$ is the refractive index of BP at the probe wavelength. Since the sound velocity is inherent of the coherent acoustic phonon and the refractive index of BP is anisotropic, the oscillation frequency is supposed to be sensitive to the probe polarization. Then, we got the oscillation frequencies $f$ = 42.51 GHz and 43.44 GHz for beam polarization along armchair and zigzag direction via the equation $1/f = \lambda/2nv$, respectively, which was consistent with the fitting results in Table 1 and the fast Fourier transform results in the inset of Figure 3d.

Furthermore, it is remarkably noted that the coherent acoustic phonon oscillation reappeared at pump-probe delay of ~185 ps as labeled by the black arrow in Figure 4. This phenomenon could be attributed to the "acoustic echo".[37-39] When the BP flake was excited by the pump pulse, the acoustic phonon wave would generate and travel from the surface to the substrate. As the wave reflected back from the substrate and arrived the surface of the BP flake, it coupled with the delayed probe pulse. Therefore, the appearance time of "acoustic echo" should be in accord with a round trip time of the acoustic phonon wave. From the thickness of the BP flake (~500 nm) and the appearance time (~185 ps) of the "acoustic echo", we could obtain the sound velocity of $5.41 \times 10^3$ m/s for the acoustic phonon in BP, which is consistent with the sound velocity obtained from the fast Fourier transform in the inset of Figure 3d.

Now we turn to examine the generation mechanism of the coherent acoustic phonons in our experiment. The acoustic phonon oscillation could be excited in two distinct mechanisms.[40-41, 60] First, the pump pulse modulates the dielectric constant directly through the deformation potential. Second, the pump pulse induces lattice heating that converts into strain and modulates the dielectric constant indirectly. In order to examine which is the dominant mechanism, we did the pump power dependent study of the transient absorption dynamics. Figure 5a and 5c shows the transient absorption dynamics for beam polarization along armchair and zigzag direction, respectively, at pump power of 50, 70, 90 and 110 μW. We extracted the initial oscillation amplitude of the coherent acoustic phonon oscillation and plotted in Figure 5b and 5d. The initial oscillation amplitude was $0.078\times10^{-6}$, $0.099\times10^{-6}$, $0.167\times10^{-6}$ and $0.193\times10^{-6}$ for pump power of 50, 70, 90 and 110 μW, respectively, when the pump pulse was polarized along armchair direction. The initial oscillation amplitude was $0.037\times10^{-6}$, $0.058\times10^{-6}$, $0.079\times10^{-6}$ and $0.094\times10^{-6}$ for pump power of 50, 70, 90 and 110 μW, respectively, when the pump pulse was polarized along zigzag direction. It can be seen that the initial oscillation amplitude for beam polarization along armchair direction was bigger than that along zigzag direction at each pump power. In addition, the initial amplitude of coherent acoustic phonons oscillations was linear with the pump power within the permissible range of error. We inferred that the direct deformation potential mechanism dominated the generation of coherent acoustic phonons in our experiment,[39] although both direct and indirect mechanism contribute to the generation of acoustic phonon.

We further analyzed the phase of the oscillation, which has distinct relations with sample dilation in direct mechanism and lattice heating in indirect mechanism.[61-62] The phase of oscillation generated from direct mechanism $\theta_D$ could be expressed by $\theta_D = \arctan(\omega_c/(1/\tau_2-1/\tau_c))$, while the phase of oscillation generated from indirect mechanism $\theta_I$ could be expressed by $\theta_I = \arctan(1/\omega_c\tau_c+\omega_c/(1/\tau_2-2/\tau_c))$.[61] Inserting the fitting results in Table 1, we could obtain $\theta_D = 47.24°$ and $\theta_I = 69.54°$, respectively, for beam polarization along armchair direction. It can be seen that $\theta_D$ is closer to the experiment result (48.32°) in Table 1 for the armchair direction. For beam polarization along zigzag direction, $\theta_D$ = 15.25° and $\theta_I$ = 25.28°, respectively. It can be seen that $\theta_D$ was much closer to the experiment result (16.14°) in the zigzag direction. This phase analysis indicates that the direct deformation potential mechanism dominated the generation of coherent acoustic phonons in our experiment, which was consistent with the amplitude analysis.

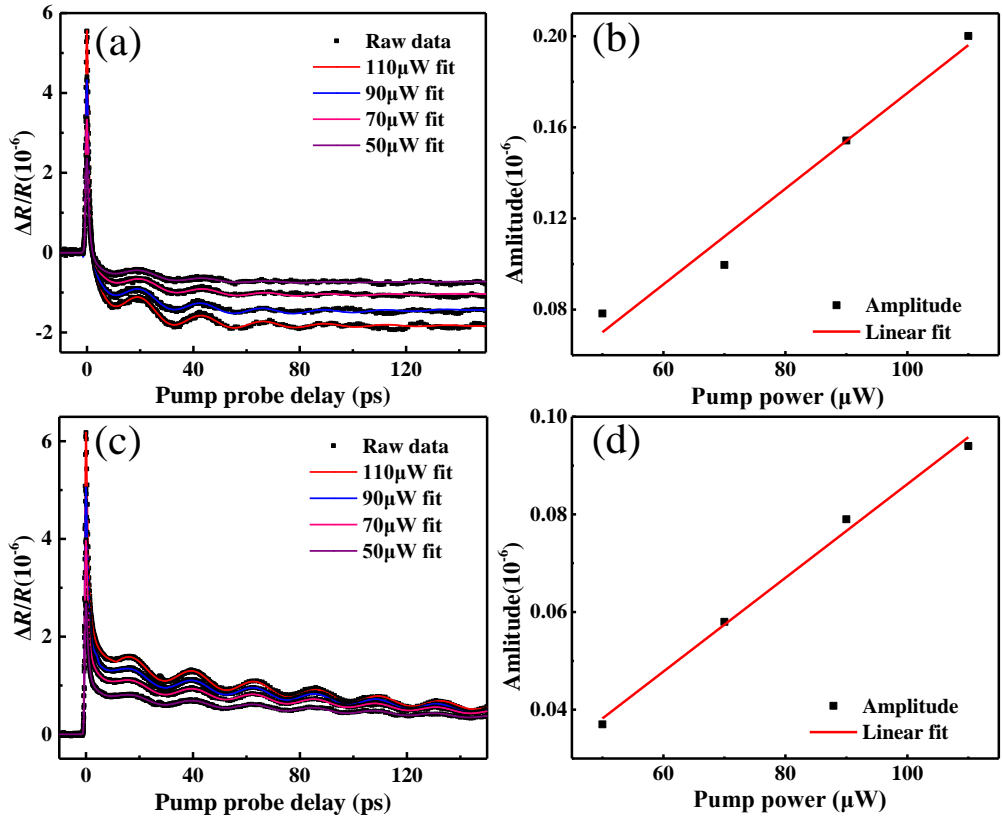

Figure 5. Transient absorption traces for pump power of 50, 70, 90 and 110 μW with beam polarization along (a) armchair and (c) zigzag directions. The black squares are raw data. The red, blue, pink and purple solid lines are the fitting results by three exponential decays superposed on one damped oscillation for pump power 110, 90, 70 and 50 μW, respectively. The probe power was fixed at 260 μW. (b) and (d) Initial amplitude of coherent acoustic phonon oscillation as a function of pump power in (a) and (c). The black squares are raw data. The red solid lines are the linear fitting results.

**Conclusions**

In this paper, the polarization-resolved transient absorption microscopy based on pump-probe technique was conducted on BP flakes in a reflection geometry to study the anisotropic dynamics of charge carriers and coherent acoustic phonon. Polarization-resolved transient absorption images and traces showed anisotropic and thickness-dependent charge carriers decay dynamics. Furthermore, anisotropic coherent acoustic phonon oscillation was observed upon the charge carriers decay. The damping time of the coherent acoustic phonon oscillation was increased when the probe polarization was changed from armchair to zigzag direction. Besides, the initial oscillation amplitude of coherent acoustic phonon oscillation was larger for probe polarization along armchair direction than that along zigzag direction. The anisotropic properties of the coherent acoustic phonon oscillation were attributed to the polarization-dependent absorption length originated from intrinsic anisotropic refractive index. Based on the amplitude and phase analysis, we proposed the direct deformation potential mechanism dominated the generation of coherent acoustic phonons in our experiment. We also calculated the sound velocity of the coherent acoustic phonon from the oscillation frequency and the "acoustic echo", respectively, which agreed well with each other. These findings provide significant insights into the rich acoustic phonon properties of BP, and promise important application for BP in polarization-sensitive optical and optoelectronic devices.

**ASSOCIATED CONTENT**

**Supporting Information**

The Supporting Information is available free of charge on the ACS Publications website at http://pubs.acs.org. Polarized Raman spectra; Transient absorption images obtained at 10 ps delay time, 0°, 45° and 90° polarization; Fitting results of transient absorption traces with polarization along armchair and zigzag direction in Figure 3.

**AUTHOR INFORMATION**

**Corresponding author**

*Email: gaobo@hit.edu.cn

**Author contributions**

B. G. initiated the idea. S. M. and H. S. did the transient absorption experiment. H. J. did the Raman experiment. S. M., H. S., H. J., X. S. and B. G. analyzed data, interpreted data, S. M. and B. G. wrote the manuscript. B. G. supervised the project.

**Notes**

The authors declare no competing financial interest.

**ACKNOWLEDGEMENT**

This work was financially supported by the National Natural Science Foundation of China (No. 21473046 and 21203046)

**TOC Graphic**

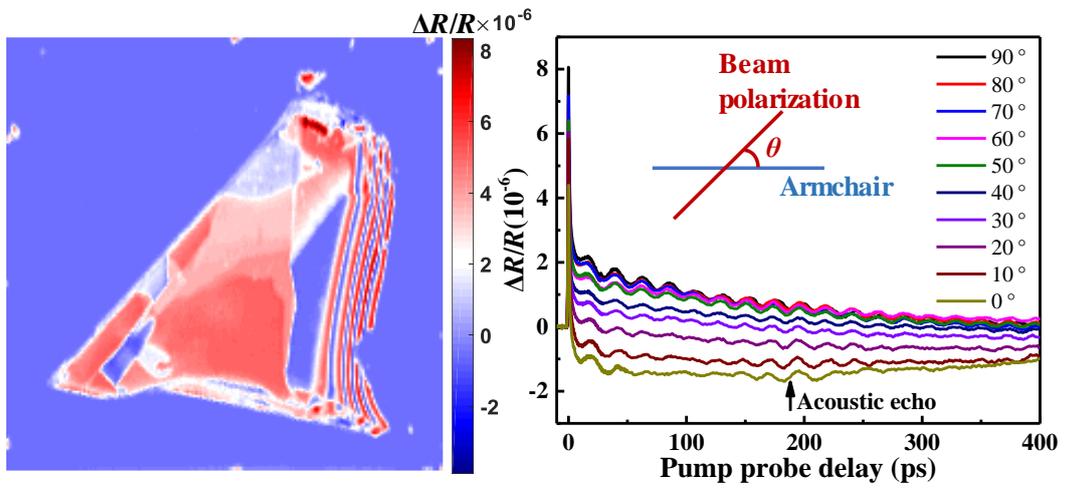